\documentclass[pra,preprint,groupedaddress]{revtex4-1}   
\usepackage[latin9]{inputenc}
\usepackage{color}
\usepackage{amsmath}
\usepackage{amssymb}
\usepackage{graphicx}

\usepackage{times}  

\makeatletter
\@ifundefined{textcolor}{}
{%
 \definecolor{BLACK}{gray}{0}
 \definecolor{WHITE}{gray}{1}
 \definecolor{RED}{rgb}{1,0,0}
 \definecolor{GREEN}{rgb}{0,1,0}
 \definecolor{BLUE}{rgb}{0,0,1}
 \definecolor{CYAN}{cmyk}{1,0,0,0}
 \definecolor{MAGENTA}{cmyk}{0,1,0,0}
 \definecolor{YELLOW}{cmyk}{0,0,1,0}
}

\newcommand{\ket}[1]{{{|}{#1}\rangle}}
\newcommand{\bra}[1]{{\langle{#1}{|}}}

\makeatother

\begin{document}

\title{Dissipative production of a maximally entangled steady state}

\author{Y. Lin}
\altaffiliation{These authors contributed equally to this work.}
\affiliation{National Institute of Standards and Technology, 325 Broadway, Boulder, CO 80305, USA}

\author{J. P. Gaebler}
\altaffiliation{These authors contributed equally to this work.}
\affiliation{National Institute of Standards and Technology, 325 Broadway, Boulder, CO 80305, USA}

\author{F. Reiter}
\affiliation{QUANTOP, The Niels Bohr Institute, University of Copenhagen, Blegdamsvej 17, DK-2100 Copenhagen $\O$, Denmark}

\author{T. R. Tan}
\affiliation{National Institute of Standards and Technology, 325 Broadway, Boulder, CO 80305, USA}

\author{R. Bowler}
\affiliation{National Institute of Standards and Technology, 325 Broadway, Boulder, CO 80305, USA}

\author{A. S. S$\o$rensen}
\affiliation{QUANTOP, The Niels Bohr Institute, University of Copenhagen, Blegdamsvej 17, DK-2100 Copenhagen $\O$, Denmark}

\author{D. Leibfried}
\affiliation{National Institute of Standards and Technology, 325 Broadway, Boulder, CO 80305, USA}

\author{D. J. Wineland}
\affiliation{National Institute of Standards and Technology, 325 Broadway, Boulder, CO 80305, USA}

\maketitle

\textbf{ Entangled states are a key resource in fundamental quantum physics, quantum cryptography, and quantum computation \cite{NC}. To date, controlled unitary interactions applied to a quantum system, so-called ``quantum gates'', have been the most widely used method to deterministically create entanglement \cite{Ladd10}. These processes require high-fidelity state preparation as well as minimizing the decoherence that inevitably arises from coupling between the system and the environment and imperfect control of the system parameters. Here, on the contrary, we combine unitary processes with engineered dissipation to deterministically produce and stabilize an approximate Bell state of two trapped-ion qubits independent of their initial state. While previous works along this line involved the application of sequences of multiple time-dependent gates \cite{Barreiro} or generated entanglement of atomic ensembles dissipatively but relied on a measurement record for steady-state entanglement \cite{Krauter}, we implement the process in a continuous time-independent fashion, analogous to optical pumping of atomic states. By continuously driving the system towards steady-state, the entanglement is stabilized even in the presence of experimental noise and decoherence.  Our demonstration of an entangled steady state of two qubits represents a step towards dissipative state engineering, dissipative quantum computation, and dissipative phase transitions \cite{Kraus, VWC, Diehl}.   Following this approach, engineered coupling to the environment may be applied to a broad range of experimental systems to achieve desired quantum dynamics or steady states. Indeed, concurrently with this work, an entangled steady state of two superconducting qubits was demonstrated using dissipation \cite{Shankar2013}.}

%


Trapped ions are one of the leading experimental platforms for quantum information processing. Here, advanced protocols using unitary quantum gates have been demonstrated, see for example Refs. \cite{Lanyon,Hanneke}. However, decoherence and dissipation from coupling to the environment remains a challenge. One approach to overcome this relies on active feedback \cite{Sayrin2011,Vijay2012,Riste2012, Brakhane2012, Schindler2013, Campagne-Ibarcq2013, Riste2013a}. Such feedback techniques may be extended to quantum error correction, which can stabilize entangled states or realize fault-tolerant quantum computations. This will, however, require high- fidelity quantum gates and large qubit overheads that are beyond the reach of current experiments \cite{Ladd10}. Recently, a complementary approach has been proposed to create entangled states or perform quantum computing by engineering the continuous interaction of the system with its environment \cite{Poyatos,Plenio,Clark,Parkins,Diehl,Kraus,  VWC,    Cavity1,  Bermudez13,Superconductor1,Superconductor2, Cormick}. In our experiment, we take a step towards harnessing dissipation for quantum information processing by producing an entangled state that is inherently stabilized against decoherence by the applied interactions in a setting fully compatible with quantum computation. With this technique, we realize maximally entangled steady states with a fidelity of $F = 0.75(3)$ by simultaneously applying a combination of time-independent fields. We also demonstrate that a stepwise application of these fields can speed up the dynamics of the scheme and achieve a fidelity of $F = 0.89(2)$ after approximately 30 repetitions. In both cases, the errors can be attributed to known experimental imperfections.

\begin{figure}
  \centering
\includegraphics[width=7 cm]{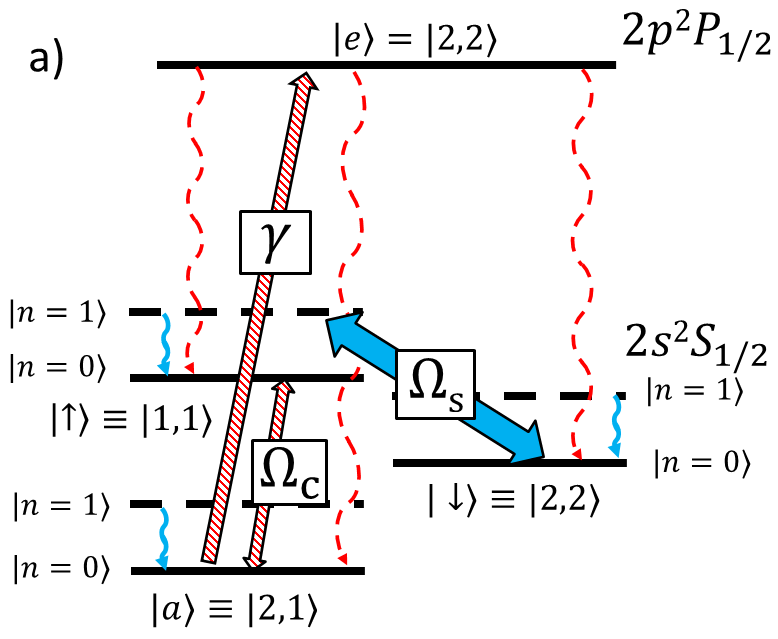}
\includegraphics[width=7 cm]{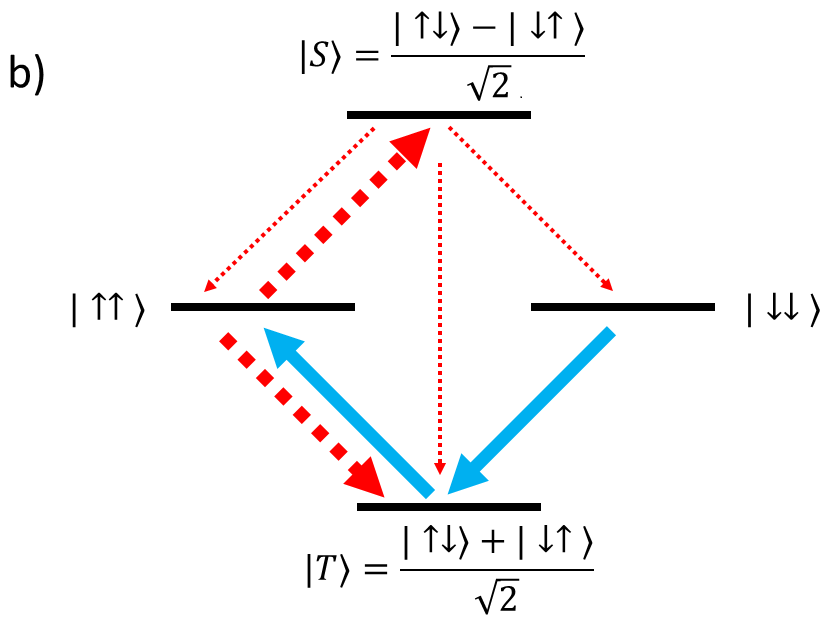} 
  \caption{\textbf{Energy Levels and Entanglement Preparation Scheme} a) The internal energy levels (not to scale) of $^9$Be$^+$ are shown as solid black lines for the ground motional state and dashed lines for the first excited motional state. The couplings needed to produce steady state entanglement are shown with blue arrows for the strong sideband coupling and sympathetic cooling and the patterned and dashed red arrows for the weak microwave coupling, repumper, and spontaneous emission from the $|e\rangle$ state.  Wavy arrows depict the dissipative processes. b) Four spin states that span the $|\uparrow\rangle$, $|\downarrow\rangle$ qubit manifold of the two $^9$Be$^+$ ions are shown as horizontal lines. Transfer processes that are accomplished by the sideband drive and sympathetic cooling are shown as blue arrows, while processes that occur by coupling the $|\uparrow\rangle$ state to the auxiliary $|a\rangle$ state followed by excitation with the repumper and decay by spontaneous emission are shown as dashed red arrows.  Processes shown as thin lines are shifted out of resonance due to the strong sideband coupling, leading to accumulation of population in the maximally entangled state $|S\rangle$ in steady state.  Further details on the rates for each process are given in the supplementary material.}\label{fig:Levels}
\end{figure}

Our scheme utilizes an ion chain with two qubit ions and at least one ``coolant" ion for sympathetic cooling \cite{Barrett03} of the qubit ions' motion. We consider a normal motional mode of this ion chain having frequency $\nu$ and mean motional quanta $\bar{n}$.  We cool the motional mode to $\bar{n}\approx0$ by laser cooling the coolant ion (or ions) and thus effectively couple the vibration to a zero-temperature bath with the phonon-loss rate denoted by ${\kappa}$. As depicted in Fig. \ref{fig:Levels}, we consider four energy levels of each qubit ion ($^9$Be$^+$), where $|\uparrow\rangle$ and $|\downarrow\rangle$ are the qubit ``spin" states,  $\ket{\rm a}$ is an auxiliary state, and $|e\rangle$ is a fast-decaying excited electronic state. A sideband excitation, with Hamiltonian $H_{s}\equiv\Omega_s(|\uparrow\rangle_{1}\langle\downarrow|+|\uparrow\rangle_{2}\langle\downarrow|)b^{+}+h.c.$ in the atomic and motional rotating frame, couples the two ions' spins via the motion, where $\Omega_s$ denotes the Rabi frequency, $b^{+}$ is the motional-mode Fock-state creation operator, the number subscripts denote the qubit ion number, and $h.c.$ is the Hermitian conjugate. A carrier interaction with Hamiltonian $H_{c}\equiv\Omega_{c}(\ket{\rm a}_{1}\langle\uparrow|+\ket{\rm a}_{2}\langle\uparrow|)+h.c.$ drives the $|\uparrow\rangle$$\leftrightarrow$ $\ket{\rm a}$ transition on each ion with Rabi frequency $\Omega_{c}$, and a repump laser incoherently drives $\ket{\rm a}$ $\mapsto$ $|\downarrow\rangle$, $|\uparrow\rangle$ by coupling to the intermediate state $|e\rangle$ with a rate of $\gamma$. All the above transitions are homogeneously driven on both qubit ions, such that individual addressing is not needed for this scheme.
These couplings ensure that the maximally entangled singlet state $|\rm S\rangle \equiv\textstyle{\frac{1}{\sqrt{2}}}(|\uparrow \downarrow \rangle - |\downarrow\uparrow \rangle)$ is the only steady state of the effective dynamics \cite{EffectiveOperators} in the regime $\gamma,\kappa,\Omega_c \ll \Omega_s$.

For an intuitive understanding of the scheme, we first consider only the sideband excitation and the sympathetic cooling (blue lines in Fig. \ref{fig:Levels} a), which, when applied together, have two dark states that are not affected by the interactions $ |\uparrow\uparrow\rangle|0\rangle$ and $\ket{\rm S}|0\rangle$. The remaining basis states of the qubits, $|\downarrow\downarrow\rangle$ and ${\ket{\rm T}\equiv\textstyle{\frac{1}{\sqrt{2}}}(|\uparrow\downarrow\rangle + |\downarrow\uparrow\rangle)}$, are driven by $H_s$ and eventually pumped to $|\uparrow\uparrow\rangle|0\rangle$ by the combination of the sideband drive and the sympathetic cooling (Fig. \ref{fig:Levels} b). The effect of adding the carrier drive $H_c$ is to couple the $|\uparrow\uparrow\rangle$ state to a combination of the $\ket{\rm \uparrow a}$, $\ket{\rm a\uparrow}$, and $\ket{\rm aa}$ states
and the $\ket{\rm S}$ state to the $\ket{\rm S_a}\equiv\frac{1}{\sqrt{2}}(\ket{\rm a\downarrow}-\ket{\rm \downarrow a})$ state. However, assuming the ions are in the ground state of motion, the dressed states of the sideband Hamiltonian $H_s$ containing $\ket{\rm S_{a}}$ have eigenenergies $\pm\Omega_s$, while $\ket{\rm S}$, $|\uparrow \uparrow\rangle$, $\ket{\rm \uparrow
a}$, and $\ket{\rm a\uparrow}$ are dark states of $H_s$ with zero eigenenergy. Thus, the transition from $\ket{\rm S}\ket{0}$ to $\ket{\rm S_a}\ket{0}$ is shifted out of resonance with the carrier drive and therefore suppressed for $\Omega_c \ll \Omega_s$. On the other hand, the transitions from the $|\uparrow\uparrow\rangle\ket{0}$ state to the $\ket{\rm \uparrow a}\ket{0}$ and $\ket{\rm a\uparrow}\ket{0}$ states are not energy shifted and remain resonant. The repumper incoherently transfers the state $\ket{\rm a}$ back to the $|\uparrow\rangle$ and $|\downarrow\rangle$ qubit manifold.  Thus, the combination of $H_c$ and the repumper create a process to pump $|\uparrow\uparrow\rangle$ to $\ket{\rm S}$ as well as a depumping process from $\ket{\rm S}$ to $|\downarrow\downarrow\rangle$, $|T\rangle$, and $|\uparrow\uparrow\rangle$, although the latter is significantly slower (Fig. \ref{fig:Levels} b).  In the limit where the rate to pump other states into $\ket{\rm S}$ is much greater than the depumping rate from $\ket{\rm S}$, the steady state will approach $\ket{\rm S}$.  The ratio of these rates can be made arbitrarily high by reducing the values of $\gamma,\kappa$ and $\Omega_c$ compared to $\Omega_s$ and in steady state the fidelity of the maximally entangled state $\ket{\rm S}$ can approach unity (see supplementary material). 

For our experimental implementation we confine a $^{9}\mathrm{Be^{+}}$-$^{24}\mathrm{Mg^{+}}$-$^{24}\mathrm{Mg^{+}}$-$^{9}\mathrm{Be^{+}}$
four-ion chain in a linear radio-frequency Paul trap described in \cite{Jost09, Hanneke}.  The two $^{9}\mathrm{Be^{+}}$ ions serve as qubit ions while the two $^{24}\mathrm{Mg^{+}}$ ions are used for sympathetic cooling.
The ion chain lies along the axis of the trap, the axis of weakest confinement, and has an extent of approximately $11$ $\mu$m. We label the four-ion axial modes \{$1,2,3,4$\},
which have mode frequencies $\nu_{1-4} \simeq$ $\{2.0,4.1,5.5,5.8\}$ MHz, respectively. An internal-state
quantization magnetic field B $\simeq$ 11.964 mT is applied along a direction $45^{\circ}$
to the trap axis, which breaks the degeneracy
of the magnetic sub-levels of $^{9}$Be$^{+}$ and $^{24}$Mg$^{+}$.
As depicted in Fig. \ref{fig:Levels} a), we utilize the $^{9}$Be$^{+}$ internal states $|F=1,m_{F}=1\rangle\equiv|\uparrow\rangle$,
$|2,2\rangle\equiv|\downarrow\rangle$, and $|2,1\rangle\equiv\ket{\rm a}$. To create the sideband coupling term $H_s$ we apply two $313$ nm laser beams in a Raman configuration tuned approximately $270$ GHz below the {$2s\ ^{2}$S$_{1/2}$ to $2p\ ^{2}$P$_{1/2}$} transition with a frequency difference equal to $f_0 + \nu_3$ where $f_0 \simeq 1.018$ GHz is the resonant transition frequency between the $|\downarrow\rangle$ and $|\uparrow\rangle$ states. The two beams are derived from the same laser and frequency-shifted using acousto-optic modulators \cite{Monroe95a}. The difference wave vector of the two beams is parallel to the trap
axis.
Microwaves are used to drive resonant transitions between the $|\uparrow\rangle$ state and the $\ket{\rm a}$ state ($f \simeq 1.121$ GHz) to create $H_c$. We also apply a repump laser beam to drive the
$\ket{\rm a}$ state to the $2p\ ^{2}P_{1/2}|2,2\rangle$ state, which subsequently
spontaneously emits a photon and decays to $|\uparrow\rangle$, $|\downarrow\rangle$
or $\ket{\rm a}$ with a branching ratio of approximately 5:4:3.  Phonon excitations due to the photon recoil are removed by the sympathetic cooling. To cool the $^{24}$Mg$^{+}$ ions, a Doppler cooling beam, two Raman-sideband beams, and a repump beam co-propagate with the $^{9}$Be$^{+}$ Raman beams. These beams ($\lambda \simeq 280$ nm) interact negligibly with the internal states of the $^{9}$Be$^{+}$ ions.  We initialize each experiment by first applying Doppler cooling to $^{9}$Be$^{+}$ and $^{24}$Mg$^{+}$, followed by $^{24}$Mg$^{+}$ sideband cooling of all the axial modes to near the ground state of motion \cite{Jost09}. An optical pumping pulse initializes the $^{9}$Be$^{+}$ ions to the $|\downarrow\downarrow\rangle$ state. We then apply the dissipative
entanglement preparation operations, as detailed below. Finally, we perform spin-state analysis to measure the populations of the $\ket{\rm S}, \ket{\rm T}, |\uparrow\uparrow\rangle,$ and $|\downarrow\downarrow\rangle$ spin states (see Methods).

%

%
\begin{figure}
  \centering
\includegraphics[width=9 cm]{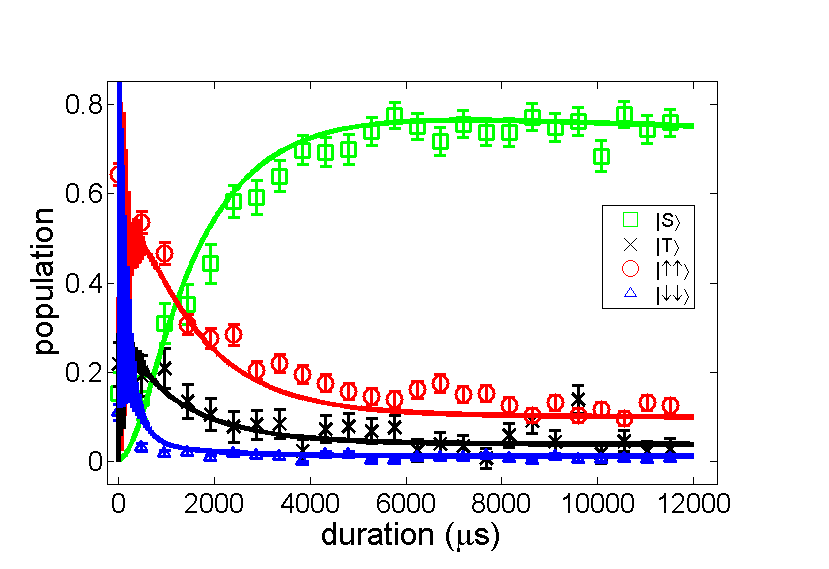}
  \caption{\textbf{Steady State Entanglement} The measured populations of the singlet, triplet, $|\uparrow\uparrow\rangle$ and $|\downarrow\downarrow\rangle$ states are shown as squares, crosses, circles, and triangles, respectively, as a function of the duration that all the elements of the dissipative entanglement scheme are applied simultaneously.  The system reaches a steady state with a $0.75(3)$ population in the target singlet state after a few ms. The solid lines are the result of a simulation based on the experimental parameters (see Methods).  The slow decrease in the singlet state fidelity at long times is due to a leak of the qubits to spin states outside the $|\uparrow\rangle$, $|\downarrow\rangle$, $\ket{\rm a}$ manifold caused by spontaneous emission from the lasers that generate the sideband coupling (see Methods and Supplementary Material). Error bars represent standard deviations of each point.}\label{fig:Fidelitiescont}
\end{figure}

\begin{figure}
  \centering
  \includegraphics[width = 9 cm]{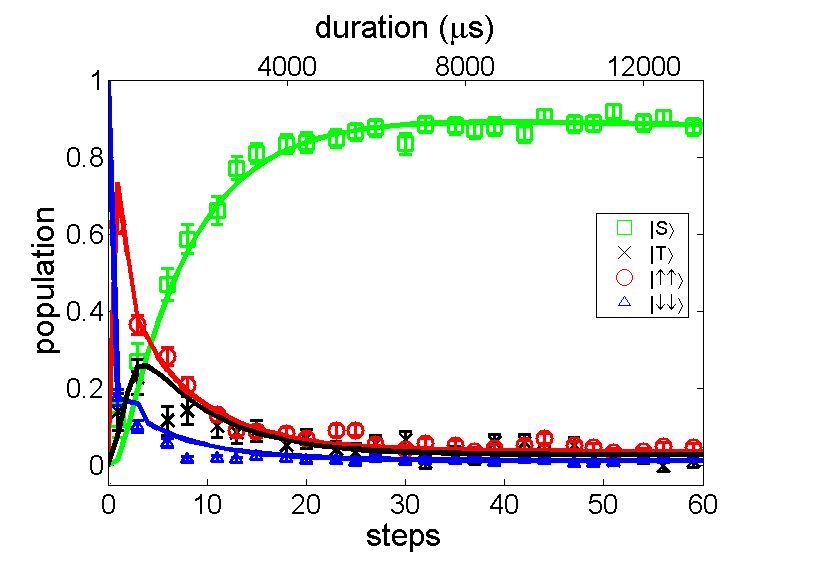}\\
  \caption{\textbf{Entanglement With Stepwise Scheme} The measured populations of the singlet, triplet, $|\uparrow\uparrow\rangle$ and $|\downarrow\downarrow\rangle$ states are shown as squares, crosses, circles, and triangles, respectively, as a function of the number of applied steps.  Each step has a duration of approximately $220$ $\mu$s.  The solid lines are the result of a model as explained in the Methods section. Error bars represent standard deviations of each point.}\label{fig:Fidelities}
\end{figure}

We implement the entanglement scheme using mode 3, where the $^{9}$Be$^{+}$ ions oscillate in phase with each other but out of phase with the $^{24}$Mg$^{+}$ ions (which oscillate in phase). In one implementation of the experiment, we apply the laser-induced sideband excitation, microwave-induced carrier excitation, repumper, and sympathetic cooling simultaneously (see methods for parameter values) for a duration $t$ and obtain a steady-state singlet state fidelity of 0.75(3), as shown in Fig. \ref{fig:Fidelitiescont}.

We model the experiment (solid lines in Fig. \ref{fig:Fidelitiescont}) taking into account: (1) the additional spontaneous emission due to the off-resonant $^9$Be$^+$ sideband laser beams, (2) the position fluctuations of those beams at the ions' location, which leads to unequal sideband Rabi rates on the two $^9$Be$^+$ ions, (3) off-resonant coupling of the sideband excitation to other motional modes, and (4) heating processes (see Methods). The model is in close agreement with the data and suggests that the dominant errors come from the spontaneous emission induced by the sideband laser beams and unequal sideband Rabi rates. In the supplementary material, we explain these errors and discuss how they can be reduced.

We also implement the scheme in a stepwise manner.  In this case we can take advantage of coherences to speed up the entanglement creation process and thereby reduce the effect of the spontaneous emission induced by the $^{9}\mathrm{Be^{+}}$ sideband laser beams. Specifically, we apply a sequence of steps with each step consisting of a coherent pulse with $H_{\mathrm{coh}} = H_s+H_c$ followed by the dissipative processes of repumping and sympathetic cooling, applied sequentially (the order does not matter). In the steady-state entanglement procedure outlined above we required $\Omega_c,\gamma, \kappa \ll \Omega_s$ to suppress transitions from $\ket{\rm S}$ to $\ket{\rm S_a}$.
However, when $H_{\mathrm{coh}}$ is applied without any dissipation, ions initially in the $\ket{\rm S}$ state will oscillate between $\ket{\rm S}$ and a superposition of $\ket{\rm S}$ and $\ket{\rm S_a}$, which is dressed by $H_s$, with a period of $2\pi/\sqrt{\Omega_s^2+\Omega_c^2}$, assuming the ions are in the motional ground state. Thus, by applying $H_{\mathrm{coh}}$ for a full oscillation period the interaction will be an identity operation for the $\ket{\rm S}$ state while all other states will be partially transferred to the auxiliary level $\ket{\rm a}$, which can then be repumped to create $\ket{\rm S}$.  However, if $n \neq 0$ some population will be transferred out of the $\ket{\rm S}$ state since the oscillation period is dependent on $n$. By taking advantage of the coherent evolution, we relax the requirement $\Omega_c,\gamma,\kappa \ll \Omega_s$ and the entanglement preparation time scale can be shortened, which reduces the error due to spontaneous emission induced by the sideband laser beams.  During the coherent process the entangled state $\ket{\rm S}$ is no longer strictly a steady state; however, if the ratio $\Omega_c/\Omega_s$ is small, the evolution of the state away from $\ket{\rm S}$ will be correspondingly small and $\ket{\rm S}$ remains an approximate steady state.


The results of the stepwise experiment are shown in Fig. \ref{fig:Fidelities}. We obtain the singlet state with fidelity $0.89(2)$.  We use the same model as for the continuous case to predict the outcome of the stepwise scheme, and find good agreement with the data (solid lines in Fig. \ref{fig:Fidelities}) with the largest sources of error coming from heating processes, unequal sideband Rabi rates, spontaneous emission caused by the $^{9}\mathrm{Be^{+}}$ sideband lasers, and off-resonant coupling of the sideband to mode 4.

In conclusion, we have presented deterministic steady state pumping into a maximally entangled state with fidelities that are limited by known experimental imperfections.  
This result can be extended to other systems where two-qubit quantum logic gates may not be feasible due to strong dissipation \cite{Cavity1}, and represents a step towards harnessing dissipation for quantum information processing.

\textbf{METHODS SUMMARY}{ 
The Methods section includes (1) the state detection and analysis procedure, (2) the experimental parameters for continuous and stepwise implementation of the scheme, and (3) the theoretical model used to produce the solid lines in Figs. \ref{fig:Fidelitiescont} and \ref{fig:Fidelities}.
}




\textbf{Acknowledgements}{  This work was supported by IARPA through ARO (Grant No. DNI-11523A1), ONR, the NIST Quantum Information Program, the European Union's Seventh Framework Program through SIQS (grant no. 600645) and through the ERC grant QIOS (grant no. 306576). We thank David Allcock and Brian Sawyer for comments on the manuscript. FR acknowledges Ben Lanyon, Rainer Blatt, and Jonathan Home for helpful conversations and acknowledges support from the Studienstiftung des deutschen Volkes. This paper is a contribution of the National Institute of Standards and Technology and is not subject to US copyright.
}

\textbf{Author Contributions}{  YL and JPG performed the experiments, analysed the data, and developed the numerical model. FR proposed the entanglement scheme and developed the analytic rate model described in the supplementary material under the guidance of AS. TRT contributed to the numerical model and the experimental apparatus. RB contributed to the experimental apparatus.  DL and DJW directed the experiments. All authors provided important suggestions for the experiments, discussed the results, and contributed to the manuscript.
}

\textbf{Author Information}{    
The authors declare no competing financial interests. Correspondence and requests for materials should be addressed to Y. Lin at yiheng.lin@colorado.edu or F. Reiter at reiter@nbi.dk.
}




\section{METHODS}


\subsection{Spin-state fidelity measurement}
To detect the populations of the $\ket{\rm S}$, $\ket{\rm T}$, $|\uparrow\uparrow\rangle$, and $|\downarrow\downarrow\rangle$ states, we need to obtain the relevant elements of the density matrix $\rho$ describing the state of the two $^{9}$Be$^{+}$ ions during the experiment. Since each ion may be found in any of the three ground states $\ket{\uparrow},\ket{\downarrow},\ket{\rm a}$ (Fig. \ref{fig:Levels}) the density matrix has dimensions $9\times9$. The singlet-state population is given by $\frac{1}{2}(\rho_{{\uparrow\downarrow},{\uparrow\downarrow}}+ \rho_{{\downarrow\uparrow},{\downarrow\uparrow}})-Re({\rho_{{\uparrow\downarrow},{\downarrow\uparrow}}})$ and the triplet-state population is given by $\frac{1}{2}(\rho_{{\uparrow\downarrow},{\uparrow\downarrow}}+ \rho_{{\downarrow\uparrow},{\downarrow\uparrow}})+Re({\rho_{{\uparrow\downarrow},{\downarrow\uparrow}}})$. The fidelity of the target entangled state, $F$, is equal to the singlet-state population. For the steady state fidelity, we report the average fidelity measured between $6$ and $12$ ms for the continuous case and between $35$ and $59$ steps in the stepwise case.
We first measure the populations
of the $|\downarrow\rangle$ state by collecting fluorescence photons from the laser-induced
cycling transition $|\downarrow\rangle\leftrightarrow2p\ ^{2}P_{3/2}|3,3\rangle$ of both  $^{9}$Be$^{+}$ ions together. We apply this detection beam for 250 $\mu$s
and collect photon counts with a photo-multiplier tube (approximately 30 counts are registered per ion in the $|\downarrow\rangle$ state). We repeat the experiment and detection 400 times to obtain a histogram. We
fit the histogram of counts to a Poisson distribution to obtain the probabilities to measure both ions, one ion, and zero ions in the $|\downarrow\rangle$ state denoted by $P_2$, $P_1$, and $P_0$, respectively. Specifically, these probabilities are related to the density matrix as follows: $P_2 = \rho_{{\downarrow\downarrow},{\downarrow\downarrow}}$, $P_1 = \rho_{{\downarrow\uparrow},{\downarrow\uparrow}}+\rho_{{\uparrow\downarrow},{\uparrow\downarrow}}+\rho_{{\rm a\downarrow},{\rm a\downarrow}}+\rho_{{\rm \downarrow a},{\rm \downarrow a}}$, and $P_0 = \rho_{{\uparrow\uparrow},{\uparrow\uparrow}}+\rho_{{\rm a\uparrow},{\rm a\uparrow}}+\rho_{{\rm \uparrow a},{\rm \uparrow a}}+\rho_{{\rm a a},{\rm a a}}$.
 We repeat the entanglement preparation scheme and perform a microwave $\pi$ pulse on the $|\downarrow\rangle \leftrightarrow |\uparrow\rangle$ followed by the same detection procedure to obtain: $P_{2,\pi} = \rho_{{\uparrow\uparrow},{\uparrow\uparrow}}$, $P_{1,\pi} = \rho_{{\uparrow\downarrow},{\uparrow\downarrow}}+\rho_{{\downarrow\uparrow},{\downarrow\uparrow}}+\rho_{{\rm a\uparrow},{\rm a\uparrow}}+\rho_{{\rm \uparrow a},{\rm \uparrow a}}$, and $P_{0,\pi} = \rho_{{\downarrow\downarrow},{\downarrow\downarrow}}+\rho_{{\rm a\downarrow},{\rm a\downarrow}}+\rho_{{\rm \downarrow a},{\rm \downarrow a}}+\rho_{{\rm a a},{\rm a a}}$.  Thus, assuming the population of the $\ket{\rm aa}$ state is negligible (see below), we have $\rho_{{\uparrow\downarrow},{\uparrow\downarrow}}+ \rho_{{\downarrow\uparrow},{\downarrow\uparrow}}= P_1-(P_{0,\pi}-P_{2})$.   To obtain the off-diagonal elements we perform the same experiment but with  a microwave $\pi/2$ pulse on the $|\downarrow\rangle \leftrightarrow |\uparrow\rangle$ transition prior to the detection to obtain $P_{2,\frac{\pi}{2}}$, $P_{1,\frac{\pi}{2}}$, and $P_{0,\frac{\pi}{2}}$. The phase of the microwave is randomized in each experiment.  It can be shown that $Re({\rho_{{\uparrow\downarrow},{\downarrow\uparrow}}}) = -1/2+2 P_{0,\frac{\pi}{2}}+\frac{1}{2}(P_2-P_0)+\frac{1}{2}(P_{2,\pi}-P_{0,\pi})$, which gives the last piece of information needed to obtain the populations of the $\ket{\rm S}$ and $|T\rangle$ states.

Due to spontaneous Raman scattering caused by the sideband laser beams it is possible that the $^{9}$Be$^{+}$ ions can be transferred to a hyperfine state outside the ${|\uparrow\rangle, |\downarrow\rangle, \ket{\rm a}}$ manifold. However, this detection procedure does not distinguish these states from the $\ket{\rm a}$ state.  Our model predicts that the probability to find at least one ion outside the three-state manifold is at most $5\%$ for the data in Fig. \ref{fig:Fidelitiescont} and $3\%$ for the data in Fig. \ref{fig:Fidelities}.  In future experiments, this population could be brought back to the three-state manifold with additional repump beams.

To calculate the singlet fidelity above, we assumed that the probability to find both atoms outside the $|\uparrow\rangle$, $|\downarrow\rangle$ qubit manifold was negligible. For the data in figures \ref{fig:Fidelitiescont} and \ref{fig:Fidelities}  we measured the probability to find at least one ion outside the qubit manifold state, given by $P_0+P_{0,\pi}-(P_2+P_{2,\pi}$), to be $7(5)\%$  and $2(2)\%$ respectively for the steady state.  We expect the probability to find both ions outside the qubit manifold to be on order of the square of the probability to find one ion outside the qubit manifold, which is therefore small.  Furthermore, our theoretical model predicts the probability of finding both ions outside the qubit manifold to be at most $1\%$ for the continuous implementation and $0.05\%$ for the stepwise implementation.

\subsection{Experimental parameters}
For the continuous implementation of the scheme shown in Fig. \ref{fig:Fidelitiescont}, the sideband Rabi rate was $\Omega_s = 2\pi\times7.8(1)$ kHz and the microwave Rabi rate was $\Omega_c = 2\pi\times 0.543(6)$ kHz. The $1/e$ time for the repump beam to deplete the $\ket{\rm a}$ state was 88 $\mu$s. The $1/e$ time for  continuous sympathetic sideband cooling of mode three was 203 $\mu$s, determined from an exponential fit of the average Fock-state occupation number $\bar{n}$ vs. sympathetic cooling time from the initial Doppler-cooled value of $\bar{n}\approx 2.5$ to a steady-state value with cooling on of $\bar{n} = 0.11(1)$.   The continuous sympathetic cooling was achieved by applying the laser-induced Raman sideband for the $^{24}$Mg$^{+}$ ions that couples the electronic ground states $|F = \frac{1}{2},m_F = -\frac{1}{2}\rangle|n\rangle \leftrightarrow |\frac{1}{2},\frac{1}{2}\rangle|n-1\rangle$ simultaneously with a repump beam that transfers $|\frac{1}{2},\frac{1}{2}\rangle|n\rangle \rightarrow |\frac{1}{2},-\frac{1}{2}\rangle|n\rangle$.  The continuous sympathetic cooling off-resonantly cooled the other axial modes $1,2,$ and $4$ with $1/e$ times of approximately $1300$ $\mu$s, $294$ $\mu$s, and $181$ $\mu$s to thermal states with average Fock state occupation numbers of approximately $2.9$, $0.95$, and $0.12$, respectively. The Rabi rate for the $^{24}$Mg$^{+}$ sideband transition on mode three was $\approx 2\pi\times11.9$ kHz and the repumper rate was $\approx 2\pi \times 625$ kHz (corresponding to a $1/e$ repump time of $1.6$ $\mu$s). The repumper rate was made significantly stronger than the sideband rate to eliminate any coherent dynamics between the $^{24}$Mg$^{+}$ spins and ion-crystal motion.

We implemented the stepwise scheme in the following way: in each step we first sympathetically cooled
each of the modes of the $^{9}\mathrm{Be^{+}}$-$^{24}\mathrm{Mg^{+}}$-$^{24}\mathrm{Mg^{+}}$-$^{9}\mathrm{Be^{+}}$ chain with $^{24}\mathrm{Mg^{+}}$ Raman sideband cooling \cite{Monroe95}, followed by application of $H_{\mathrm{coh}}$ for a duration $t_{2\pi}$, and at the end of each step we applied the repumper. The populations of the qubit state were measured at the end of each step and plotted in Fig. \ref{fig:Fidelities}. The $^{9}$Be$^{+}$ sideband Rabi rate was $\Omega_s = 2\pi\times8.4(1)$ kHz and the microwave Rabi rate was $\Omega_c = 2\pi\times1.24(6)$ kHz.  The repumper had a $1/e$ time of approximately $3$ $\mu$s and was turned on for $6$ $\mu$s in each step.  In each step, two sympathetic cooling cycles were applied to mode 1, which has the largest heating rate, and one pulse was applied to each of the remaining modes, with mode 3 being the last. A sympathetic cooling cycle consists of a single motion subtracting sideband pulse applied to the $^{24}$Mg$^{+}$ ions followed by a repump pulse.  The duration to apply all the cooling pulses was approximately $100$ $\mu$s in each step.

In both cases the ion spacing was set by adjusting the strength of the harmonic confinement, such that $ \Delta_k z = 2 \pi m$ where $\Delta_k \approx \frac{2 \pi \sqrt{2}}{313\times10^{-9}}$ m$^{-1}$  is the wavevector difference of the $^{9}$Be$^{+}$ Raman sideband lasers, $z$ is the distance between the $^{9}$Be$^{+}$ ions, and $m$ is an integer, such that the phase of the sideband excitation was equal on both ions.  For our confinement strength, $z \simeq 11$ $\mu$m such that the value of $m$ was near $300$. For modes where the qubit ions move in phase, the integer value of $m$ ensures $H_s$ is as defined in the main text. However, in the general case $H_{s}\equiv\Omega_s(|\uparrow\rangle_{1}\langle\downarrow|+e^{i\phi}|\uparrow\rangle_{2}\langle\downarrow|)b^{+}+h.c.$, where $\phi$ is the phase difference between the two $^{9}$Be$^{+}$ ions of the sideband coupling, and the steady state of the system (including the cooling, repumper, and microwave carrier) is $|D_\phi\rangle \equiv\frac{|\uparrow\downarrow\rangle-e^{i\phi}|\downarrow\uparrow\rangle}{\sqrt{2}}$.

\subsection{Numerical model}

We modeled our experiment using a master equation with a coherent component describing the $^{9}$Be$^{+}$ sideband and microwave carrier drives and Lindblad operators describing the sympathetic cooling, repumper, and spontaneous emission due to the  $^{9}$Be$^{+}$ sideband lasers.  The coherent Hamiltonian is
\begin{align*}
H_{coh}\equiv\ &\Omega_s[(1-\frac{r}{2})|\uparrow\rangle_{1}\langle\downarrow|+(1+\frac{r}{2})|\uparrow\rangle_{2}\langle\downarrow|)]b^{+} \\
 &+\Omega_{c}(\ket{\rm a}_{1}\langle\uparrow|+\ket{\rm a}_{2}\langle\uparrow|)+h.c.,
\end{align*}
where $r$ describes the Rabi-rate imbalance of the sideband on the two ions. The Lindblad operator describing sympathetic cooling is given by $L_{\kappa} = \sqrt{\kappa}b$, and the repumper is given by $L_{\gamma_{j,\rm a}}$, where $j$ is either the $|\uparrow\rangle$ or $|\downarrow\rangle$ state and $L_{\gamma_{j,\rm a}}=\sqrt{\gamma_{j,\rm a}}|j\rangle\langle \rm a|$.  Heating processes that limit the sympathetic cooling are modeled with a Lindblad operator $L_{\kappa_h} = \sqrt{{\kappa_h}}b^{\dagger}$, where $\kappa_h$ is determined experimentally by measuring $\bar{n}$ for mode three after sympathetic cooling (no other interactions are turned on). The heating rate is given by $\kappa_h = \frac{\kappa \bar{n}}{1+\bar{n}}$.  For the continuous cooling used for the data in Fig. \ref{fig:Fidelitiescont} we found $\bar{n} = 0.11(1)$ and for the stepwise case of Fig. \ref{fig:Fidelities} we found $\bar{n} = 0.08(1)$. We take into account spontaneous emission that incoherently changes population from the state $i$ to the state $j$ $(i \neq j)$ caused by the $^{9}$Be$^{+}$ sideband laser beams with Lindblad operators of the form $L_{j,i} = \sqrt{\Gamma_{j,i}}|j\rangle\langle i|$, where $\Gamma_{j,i}$ can be calculated using the Kramers-Heisenberg formula \cite{Ozeri}. The error caused by Rayleigh scattering ($i=j$) is negligible \cite{Uys2010}. Off-resonant coupling to mode four is taken into account with an additional Hamiltonian term $H_4 = \Omega_s\frac{\eta_4}{\eta_3}(|\uparrow\rangle_{1}\langle\downarrow| -|\uparrow\rangle_{2}\langle\downarrow|)c^{+}e^{-i\delta t}+h.c.$, where $c^{+}$ is the raising operator for the fourth mode, $\delta \approx 2 \pi \times 250$ kHz is the splitting between modes three and four, and $\eta_3 = 0.180$ and $\eta_4=0.155$ are the Lamb-Dicke parameters of modes three and four, respectively.
The continuous implementation of the scheme is modeled by numerically solving a master equation that includes all terms for a variable duration and a given value of $r$.  We then obtain the theoretical prediction shown in Fig. \ref{fig:Fidelitiescont} by averaging simulations with different values of $r$ using a Gaussian distribution with an r.m.s value of $0.014$. This r.m.s. value was determined from fits to qubit Rabi flopping curves for a single $^{9}$Be$^{+}$ ion and for the $^{9}\mathrm{Be^{+}}$-$^{24}\mathrm{Mg^{+}}$-$^{24}\mathrm{Mg^{+}}$-$^{9}\mathrm{Be^{+}}$ ion chain.  Percent-level fluctuations of $\Omega_s$ cause negligible changes to the predicted fidelity. The result of the calculation at the end of each step is plotted in Fig. \ref{fig:Fidelities}. In both cases, the initial state of the $^{9}$Be$^{+}$ ions was taken to be $|\downarrow\downarrow\rangle|n=0\rangle$.  The particular initial state chosen affects the dynamics only at short times and does not affect the steady state.  All numerical models were implemented by use of the quantum optics toolbox \cite{QOtoolbox}.

\section{Supplementary Information}

The entanglement creation scheme presented here can in principle produce maximally entangled states with arbitrarily low error. Because of experimental limitations, the observed fidelities for the steady states created were, however, $0.75(3)$ for the continuous implementation of the scheme and $0.89(2)$ for the stepwise implementation. Here we examine the sources of error for the experiments and discuss the prospects for reducing these errors to achieve high-fidelity entangled states without the use of quantum gates. To this end, we utilize both a simplified rate model to approximate the dynamics of the system and a direct numerical integration of the master equation (described in the Methods Section).

\subsection{Rate Model}
\begin{figure}[h]
  \centering
  \includegraphics[width=7 cm]{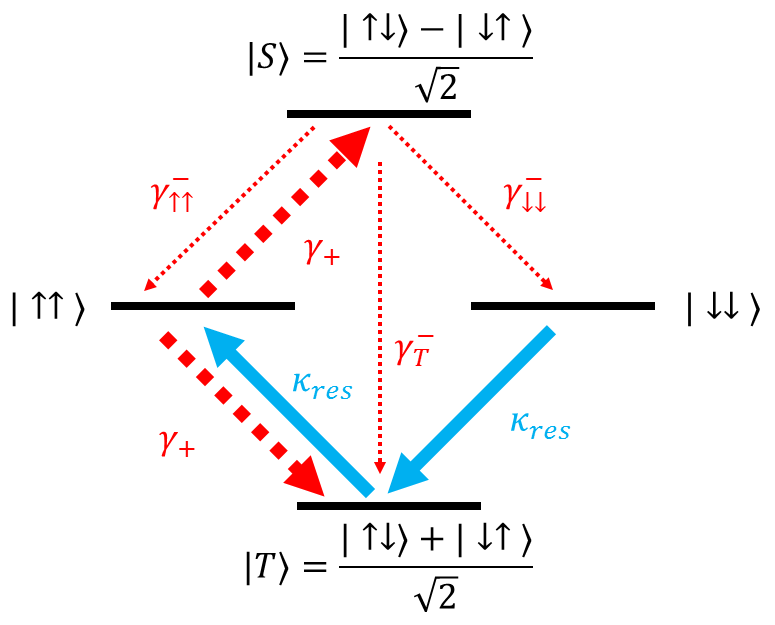}
  \caption{\textbf{Rate Diagram} The dynamics of the ground states are modelled by the rates of effective decay processes. These are the preparation rate for the singlet state $\gamma_+$, the loss rates $\gamma^-_{\uparrow \uparrow}$, $\gamma^-_{\rm T}$, and $\gamma^-_{\downarrow \downarrow}$, and the reshuffling rate $\kappa_{\rm res}$.}
\label{fig:rates}
\end{figure}

In our simplified rate model, we restrict the dynamics of the master equation to the ground states $\ket{\uparrow \uparrow}$, $\ket{\downarrow \downarrow}, \ket{\rm T}$ and $\ket{\rm S}$, due to the fast repumping of the auxiliary level. We achieve this using an effective operator formalism
\cite{EffectiveOperators} to eliminate the decaying states. Thereby we obtain effective decay processes such as the preparation rate of the singlet and loss processes from the singlet. Our model involves a rate $\gamma_+$ for the
preparation of the singlet from $\ket{\uparrow \uparrow}$. The same process also induces a decay at the same rate $\gamma_+$ from $\ket{\uparrow \uparrow}$ to $\ket{\rm T}$ since the repumper incoherently pumps each ion independently. Furthermore, the reshuffling process that transfers $\ket{\downarrow \downarrow}$ to $\ket{\rm T}$, and $\ket{\rm T}$ to $\ket{\uparrow \uparrow}$ is described by a rate $\kappa_{\rm res}$.
The losses from the singlet are modeled by three loss rates $\gamma^-_{i}$,
$i \in \{\uparrow \uparrow, \downarrow \downarrow, \rm T\}$ (overall loss rate $\gamma_- = \gamma^-_{\uparrow \uparrow} + \gamma^-_{\rm T} + \gamma^-_{\downarrow \downarrow}$), which can account for various loss processes present in the experiment. These interactions are illustrated in Fig. \ref{fig:rates}.
As no coherences between the ground states are established by these processes, the dynamics of the coherences can be dropped from the master equation. The time evolution of the ground states is then described by rate equations of their populations. With the rates introduced above these equations read
\begin{align}
\label{EqRate1}
\dot{P}_{\rm S} &= + \gamma_+ P_{\uparrow \uparrow} - (\gamma^-_{\uparrow \uparrow} + \gamma^-_{\rm T} + \gamma^-_{\downarrow \downarrow}) P_{\rm S} \\
\dot{P}_{\uparrow \uparrow} &= - 2 \gamma_+ P_{\uparrow \uparrow} + \kappa_{\rm res} P_{\rm T} + \gamma^-_{\uparrow \uparrow} P_{\rm S}\\
\dot{P}_{\rm T} &= + \gamma_+ P_{\uparrow \uparrow} - \kappa_{\rm res} P_{\rm T} + \kappa_{\rm res} P_{\downarrow \downarrow} + \gamma^-_{\rm T} P_{\rm S}\\
\dot{P}_{\rm \downarrow \downarrow} &= - \kappa_{\rm res} P_{\downarrow \downarrow} + \gamma^-_{\downarrow \downarrow} P_{\rm S}.
\label{EqRate4}
\end{align}
The effect of other decay processes acting on the triplet states (suppressed by $\Omega_{\rm c}^2/\Omega_{\rm s}^2$) is negligible compared with the fast reshuffling of these states and is thus not considered.
Setting $\dot{P}_i = 0$ for all states $i$ the steady state can then be read off from the system of coupled equations.
The fidelity of the steady state with the maximally entangled singlet is given by
\begin{align}
F = \frac{1}{1 + \mathcal{E}} \simeq 1 - \mathcal{E}
\label{EqSteady},
\end{align}
with
\begin{align}
\mathcal{E} = \frac{\gamma_-}{\gamma_+} + \frac{\gamma^-_{\rm \uparrow \uparrow} + 2 \gamma^-_{\rm T} + 3 \gamma^-_{\downarrow \downarrow}}{\kappa_{\rm res}}.
\label{EqError}
\end{align}
The error of the protocol, i.e., the infidelity of the steady state with the singlet state, $\mathcal{E} \simeq 1-F$, is thus determined by the ratios of the depumping rates out of the singlet and the pumping rates of other states into the singlet, which result in a steady-state population of the three triplet states. Therefore, processes that affect either of these rates can cause error.
Eq. (\ref{EqError}) contains two types or errors:
The first term accounts for the ratio between preparation of and loss from the singlet state and equals the error for the case of perfect reshuffling ($\kappa_{\rm res} \rightarrow \infty$) of the triplet states. The second term reflects the need to reshuffle population lost from the singlet state to $\ket{\downarrow \downarrow}$ and $\ket{\rm T}$ to $\ket{\uparrow \uparrow}$ in order to transfer it to the singlet again.
In the following, we use this model as a framework to include the rates of the desired, engineered decay processes, as well as the experimental sources of loss. We thereby obtain a simple quantitative model for the dynamics observed in the experiment.

\subsection{Entanglement Preparation}
We start out with the entanglement preparation process:
The rate for pumping other states to the singlet state is dependent on the process that takes $\ket{\uparrow\uparrow}\ket{n=0}$ to $\ket{\rm S}\ket{0}$, which is achieved by weak excitation from $\ket{\uparrow \uparrow}$ to $\ket{\rm T_{\rm a}} = \frac{1}{\sqrt{2}}(\ket{\rm a \uparrow} + \ket{\rm \uparrow a})$ and subsequent decay into $\ket{\rm S}$. For weak microwave driving this results in an effective decay by spontaneous emission from $\ket{\uparrow \uparrow}$ to $\ket{\rm S}$ with a rate \cite{EffectiveOperators}
\begin{align}
\gamma_+ = \frac{4 \gamma_{\rm \downarrow a} \Omega_{\rm c}^2}{\gamma^2},
\label{EqPrep}
\end{align}
where $\Omega_{\rm c}$ denotes the microwave carrier Rabi rate, the repumper rates are $\gamma_{\rm \downarrow a}$ (for repumping from $\ket{\rm a}$ to $\ket{\downarrow}$) and $\gamma_{\rm \uparrow a}$ (for repumping from $\ket{\rm a}$ to $\ket{\uparrow}$). Here, all decay rates are written as $\gamma_{ij}$, leading to a state $i$ from a state $j$. The line width of level $\ket{\rm a}$ is given by $\gamma = \gamma_{\rm \downarrow a} + \gamma_{\rm \uparrow a} + \gamma_{\rm aa}$. The same process transfers population from $\ket{\uparrow \uparrow}$ to $\ket{\rm T}$ at the same rate $\gamma_+$.

Once the drive $\Omega_{\rm c}$ from $\ket{\uparrow \uparrow}$ to $\ket{\rm T_a}$ approaches the line width of $\ket{\rm T_a}$ ($\gamma$) the excitation is no longer overdamped and the dynamics become more coherent and the above expression becomes inaccurate. The accuracy can be restored by including power broadening and the steady population of the excited level. This results in an adjusted preparation rate
\begin{align}
\gamma_+ = \frac{4 \gamma_{\rm \downarrow a} \Omega_{\rm c}^2}{\gamma^2 + 16 \Omega_{\rm c}^2}.
\label{EqPrep2}
\end{align}
In the simulated curves below we plot the sum of the populations of the coupled states $\ket{\uparrow \uparrow}$ and $\ket{\rm T_a}$ since these are mixed by the relatively strong drive $\Omega_{\rm c}$.


The preparation process from $\ket{\uparrow \uparrow}$ to $\ket{\rm S}$ requires the ions to be in the motional ground state.  This is because the transitions from the $\ket{\uparrow \uparrow}$ state to states containing $\ket{a}$ are shifted out of resonance with the carrier drive by the sideband coupling for $\ket{n\neq0}$. Thus, imperfect cooling slows the preparation rate for $\ket{\rm S}$, which lowers the fidelity. For a nonzero population of the higher motional states, the preparation rate thus has to be multiplied by the probability to be in the motional ground state, $P_0 = \frac{1}{1 + \bar{n}}$. We thus obtain the preparation rate
\begin{align}
\gamma_+ &= \gamma_+(\bar{n}=0) P_0 = \frac{\gamma_{\rm \downarrow a} \Omega_{\rm c}^2}{\gamma^2 (1 + \bar{n})}
\end{align}
for weak driving, or
\begin{align}
\gamma_+ = \frac{\gamma_{\rm \downarrow a} \Omega_{\rm c}^2}{(\gamma^2 + 4 \Omega_{\rm c}^2)(1 + \bar{n})},
\label{EqPrep3}
\end{align}
including the strong driving effects from above.  In the continuous experiment the motional mode is cooled to about $\bar{n} = 0.1$, which leads to a decrease in the preparation rate and an error for the singlet state of approximately $0.02$ according to both the numerical simulation and the rate model. 

In order to transfer population from all states to $\ket{\rm S}\ket{0}$ via $\ket{\uparrow \uparrow}\ket{0}$, the population from $\ket{\downarrow \downarrow}\ket{0}$ is reshuffled to $\ket{\rm T}\ket{0}$ by excitation to $\ket{\rm T}\ket{1}$ through the sideband coupling and subsequent decay to $\ket{\rm T}\ket{0}$ at a rate $\kappa$. Similarly, the population of $\ket{\rm T}\ket{0}$ is transferred to $\ket{\uparrow \uparrow}\ket{0}$ through $\ket{\uparrow \uparrow}\ket{1}$. Given that $\Omega_{\rm s} \gg \kappa$, the population oscillates back and forth several times between the coupled states before a decay happens. We can therefore assume the population spends half of the time in the phonon-excited state (and the other half in the respective ground state). The decay rate of the reshuffling process can then be approximated as \cite{EffectiveOperators}
\begin{align}
\kappa_{\rm res} \approx \frac{\kappa}{2},
\label{EqRes}
\end{align}
regardless of the actual value of the sideband coupling $\Omega_{\rm s}$.

\subsection{Inherent Depumping}
We now turn to the loss processes:
The only depumping rate inherent to the scheme is due to the off resonant coupling of the $\ket{\rm S}$ state to the state $\ket{\rm S_{a}} = \frac{1}{\sqrt{2}}(\ket{\rm a \downarrow}-\ket{\rm \downarrow a})$ and the decay from there into various states. This process is inhibited by the energy splitting $\Omega_{\rm s}$ induced by the strong sideband driving such that the inherent depumping rate from the singlet amounts to \cite{EffectiveOperators}
\begin{align}
\gamma^-_{\rm inh} = \frac{(\gamma + \kappa) \Omega_{\rm c}^2}{4 \Omega_{\rm s}^2},
\label{eq:gamma_inherent}
\end{align}
where a fraction $\gamma_{\rm \downarrow a}/(\gamma_{\rm \uparrow a} + \gamma_{\rm \downarrow a})$ decays to $\ket{\downarrow \downarrow}$, a fraction $\frac{1}{2}\gamma_{\rm \uparrow a}/(\gamma_{\rm \uparrow a} + \gamma_{\rm \downarrow a})$ decays to $\ket{\rm T}$ and the same amount returns to $\ket{\rm S}$. We use the rate equation model to quantify this source of error. For the parameters of the experiment we find that for the continuous operation an error of about $0.11$ originates from the inherent loss processes. These loss processes are not present in the stepwise scheme since (1) the repumper is applied separately from the coherent drives and (2) we adjust the duration of the coherent pulse such that all population has returned to the singlet state at its end.

The inherent loss rate $\gamma^-_{\rm inh}$ derived above can be decreased by increasing the sideband coupling $\Omega_{\rm s}$, and there is thus no fundamental limitation to the achievable fidelity of the scheme, which can ideally approach unity. In practice, there is always a limitation to the available sideband coupling strength and the parameters of the experiments thus have to be optimized given the available sideband coupling strength. The inherent loss rate (\ref{eq:gamma_inherent}) can also be decreased by decreasing the drive $\Omega_{\rm c}$. In the experiment we are, however, limited by the $\gamma_-/\gamma_+$ term in Eq. (\ref{EqError}), and since the desirable process $\gamma_+$ also decreases with decreasing $\Omega_{\rm c}$, this will only increase the necessary waiting time to reach the steady state and will not improve the fidelity. Furthermore a small $\Omega_{\rm c}$ will also increase the effect of other experimental imperfections that cause depumping from the entangled state due to the lower preparation rate of the state. In the experiment we therefore set $\Omega_{\rm c} \approx \gamma/4$, which is the point where the desirable rate $\gamma_+$ begins to become limited by the saturation effect included in Eq. (\ref{EqPrep2}). The remaining parameters $\gamma$ and $\kappa$ of our experiment are then determined by the tradeoff between the reduction of $\gamma^-_{{\rm inh}}$ (favoring low $\gamma$ and $\kappa$) and minimization of other loss processes (favoring fast preparation through large $\gamma$ and $\kappa$).

The remaining sources of error are not inherent to the scheme but arise from the particular setup used for the implementation. These are (1) spontaneous emission caused by the $^{9}$Be$^{+}$ Raman sideband lasers, (2) fluctuations of laser and microwave powers and spatial alignments, (3) heating of the motional mode, (4) off resonant coupling of the $^{9}$Be$^{+}$ Raman sideband lasers to the carrier and other motional modes and 5) magnetic field gradients and fluctuations.

\subsection{Raman Sideband Coupling Induced Spontaneous Emission}
Because we implement the sideband coupling with a Raman laser configuration, the ions have a small amplitude in the electronically excited state from which they can spontaneously emit photons, reducing an entangled spin state to a mixed state.
This results in a decay from $\ket{\rm S}$ to $\ket{\uparrow \uparrow}$ at a rate \cite{EffectiveOperators}
\begin{align}
\Gamma^-_{\uparrow \uparrow} = &\Gamma_{\uparrow \downarrow} + \frac{\Gamma_{\rm \uparrow a} \Gamma_{\rm a \downarrow}}{\Gamma_{\rm \uparrow a} + \Gamma_{\rm \downarrow a}} + \frac{\Gamma_{\rm \uparrow a} \Gamma_{\rm a \uparrow}}{2(\Gamma_{\rm \uparrow a} + \Gamma_{\rm \downarrow a})} \times \nonumber \\ &\times \left(1 + \frac{\kappa/2}{\Gamma_{\rm \uparrow a} + \Gamma_{\rm \downarrow a} + \kappa/2}\right) ,
\end{align}
from $\ket{\rm S}$ to $\ket{\rm T}$ (as well as from $\ket{\rm S}$ to $\ket{\rm S}$) at a rate
\begin{align}
\Gamma^-_{\rm T} = &\frac{\Gamma_{\rm \downarrow a} \Gamma_{\rm a \downarrow}}{2(\Gamma_{\rm \uparrow a} + \Gamma_{\rm \downarrow a})} + \frac{(\Gamma_{\rm \uparrow a} + \Gamma_{\rm \downarrow a}) \Gamma_{\rm a \uparrow}}{4(\Gamma_{\rm \uparrow a} + \Gamma_{\rm \downarrow a} + \kappa/2)} + \nonumber \\ &+ \frac{\Gamma_{\rm \downarrow a} \Gamma_{\rm a \uparrow}}{2(\Gamma_{\rm \uparrow a} + \Gamma_{\rm \downarrow a})} \frac{\kappa/2}{\Gamma_{\rm \uparrow a} + \Gamma_{\rm \downarrow a} + \kappa/2},
\end{align}
and from $\ket{\rm S}$ to $\ket{\downarrow \downarrow}$ at a rate
\begin{align}
\Gamma^-_{\downarrow \downarrow} = \Gamma_{\downarrow \uparrow} + \frac{\Gamma_{\rm \downarrow a} \Gamma_{\rm a \uparrow}}{2 (\Gamma_{\rm \uparrow a} + \Gamma_{\rm \downarrow a} + \kappa/2)}.
\end{align}
The effect of the dephasing from Rayleigh scattering $\ket{\uparrow} \leftrightarrow \ket{\uparrow}$ and $\ket{\downarrow} \leftrightarrow \ket{\downarrow}$ is negligible \cite{Ozeri,Uys2010}. The spontaneous emission rates can be calculated with the Kramers-Heisenberg formula \cite{Ozeri,Uys2010} and are proportional to the Rabi rate of the Raman sideband coupling. However, the ratio of the spontaneous emission rates to the Rabi rate can be reduced by increasing the Raman detuning from the excited state. The Raman detuning used here was $270$ GHz below the {$2s\ ^{2}$S$_{1/2}$ to $2p\ ^{2}$P$_{1/2}$} transition and the spontaneous emission rates are on the order of $10^{-4} \times \Omega_{\rm s}$.


In addition, spontaneous emission causes loss from the state $\ket{\uparrow} = \ket{1,1}$ to the $\ket{2,0}$ and $\ket{1,0}$ states, which are not repumped. As also addressed in Methods section A), this error can result in a decrease in fidelity. The additional losses to these states can be modeled by adding
\begin{align}
\dot{P}_{\uparrow \uparrow} &= ... - 2 \Gamma_{\uparrow} P_{\uparrow \uparrow} \\
\dot{P}_{\rm T} &= ... - \Gamma_{\uparrow} P_{\rm T} \\
\dot{P}_{\rm S} &= ... - \Gamma_{\uparrow} P_{\rm S},
\end{align}
where $\Gamma_{\uparrow}$ denotes the spontaneous emission rate from $\ket{\uparrow}$ to states other than $\ket{\uparrow}$, $\ket{\downarrow}$ and $\ket{\rm a}$, and the dots represent the terms in Eqs. (\ref{EqRate1})-(\ref{EqRate4}). From the simulations we find the population of states containing at least one ion in either the $\ket{2,0}$ or $\ket{1,0}$ states is approximately $0.05$ for the continuous case (averaging between $6$ and $12$ ms) and $0.03$ for the stepwise case (averaging from $35$ to $59$ steps).  These populations will continue to increase for increasing duration of the applied fields. According to our simulation, the singlet state fidelity for the continuous case would drop to $50$\% at approximately $84$ ms. In the future, this loss could be avoided by repumping the $\ket{1,0}$ and $\ket{2,0}$ states back to the qubit states.

We have performed a numerical simulation with identical parameters to the experiment but eliminated all spontaneous emission errors (while still including all other sources of error) and find that the fidelities increase by approximately $0.07$ for the continuous and $0.04$ for the stepwise implementations of the scheme. Similar results are obtained for the rate equation mode in the continuous case.  Spontaneous emission errors could be reduced by increasing the Raman sideband detuning and correspondingly increasing the laser intensity to keep the sideband Rabi rate constant. Another potential future option would be to create the sideband coupling with microwaves, which would eliminate all spontaneous emission errors from the sideband excitation \cite{Ospelkaus2008,Ospelkaus2011}.

\subsection{Experimental Apparatus Noise}
Due to fluctuations in the intensity of the laser beams and microwave fields  (typically on the order of a percent), the values of $\Omega_{\rm s},\Omega_{\rm c},\gamma$, and $\kappa$ will vary. However, for the continuous implementation of the scheme, insensitivity to fluctuations in the parameters is inherent to the method since the pumping effect relies only on that ratios between certain parameters be small, a major asset of dissipative state preparation.  For the stepwise implementation of the scheme, however, there is a greater dependence of the fidelity on the sideband Rabi rate that arises from the coherent portion of each step. Nevertheless, in the limit $\Omega_{\rm c} \ll \Omega_{\rm s}$, the decrease in fidelity due to Rabi rate fluctuations can still be small. In our experiment we estimate $\frac{\delta\Omega_{\rm s}}{\Omega_{\rm s}} = 0.008$, where $\delta\Omega_{\rm s}$ is the r.m.s fluctuation in $\Omega_{\rm s}$, and this reduces the fidelity of the entangled state by less than $0.01$ according to our numerical simulations.


%
A more significant problem for the scheme is fluctuations in the position of the $^{9}$Be$^{+}$ Raman sideband laser beams at the site of the ions. Because the lasers are each aligned at $45^{\circ}$ to the crystal axis, fluctuations in the beam positions cause unequal Rabi rates on the two $^{9}$Be$^{+}$ ions. As above, this effect can be modeled with a modified sideband Hamiltonian $H_{s}\equiv\Omega_{\rm s}[(1-\frac{r}{2})\ket{\uparrow}_{1}\bra{\downarrow}+(1+\frac{r}{2})\ket{\uparrow}_{2}\bra{\downarrow}]b^{+}+h.c.$, where $r$ characterizes the imbalance. In our
experiment we estimate that the value of $r$ fluctuates about zero from experiment to experiment with an r.m.s. deviation of approximately $0.014$. A minor source of error caused by $r \neq 0$ is that the dark state of the system is no longer the singlet state but rather $\ket{{\rm S}_r} = \frac{1}{\sqrt{2+r^2/2}}[(1-\frac{r}{2})\ket{\uparrow\downarrow} - (1 + \frac{r}{2})\ket{\downarrow\uparrow}]$. The error from the difference between $\ket{\rm S}$ and $\ket{{\rm S}_r}$ is proportional to $r^2$, which is negligible in our case.
However, when the ions are not cooled to the ground state the above Hamiltonian creates an additional depumping process for the  $\ket{{\rm S}_r}$ state. Specifically, the state $\ket{{\rm S}_r}\ket{1}$ is coupled by the sideband coupling to  $\ket{\rm D} \equiv \frac{1}{\sqrt{3}}(\ket{\uparrow\uparrow}\ket{2} - \sqrt{2} \ket{\downarrow\downarrow}\ket{0})$ with a Rabi rate $\frac{2}{\sqrt{3}} r \Omega_{\rm s}$. With sympathetic cooling
$\ket{\rm D}$ decays towards $\ket{\uparrow \uparrow}\ket{0}$ with a rate given by $\frac{2 \kappa}{3}$. Taking into account the fraction of phonon-excited population $P_{> 0} = \frac{\bar{n}}{\bar{n}+1} \approx \bar{n}$, we find an effective decay from $\ket{{\rm S}_r}$ to $\ket{\rm \uparrow \uparrow}$ at a rate
\begin{align}
\kappa^-_r \approx \frac{16 (r \Omega_{\rm s})^2 \bar{n}}{5 \kappa}.
\end{align}
In the ideal case, with no heating processes, the steady state will be $\ket{{\rm S}_r} \approx \ket{\rm S}$ and this depumping process can be avoided. However, as discussed below, the ions are cooled only to a steady state with $\bar{n} \approx 0.1$, and this depumping process causes errors in both the continuous and stepwise experiments.
We perform a numerical simulation with identical parameters to the experiment but set $r=0$ (while still including all other sources of error) and find the fidelity increases by $0.02$ for the continuous (obtained from both the simulation and the rate equation model) and $0.01$ for the stepwise implementation of the scheme. This source of error could be reduced or eliminated in several ways. For example, stabilizing the alignment of the beams will reduce fluctuations. A better approach would be to align the Raman beams to counter-propagate along the ion crystal axis. In this case alignment fluctuations would cause only fluctuations in $\Omega_{\rm s}$ but not $r$. Potentially another approach would be to create the sideband coupling using near-field microwaves (which would also eliminate the spontaneous emission errors) \cite{Ospelkaus2008, Ospelkaus2011}.

The singlet state is insensitive to fluctuations in the magnetic field; however, gradients of the magnetic field lead to each qubit ion experiencing a different magnetic field, which breaks the degeneracy of the $\ket{\uparrow\downarrow}$ and $\ket{\downarrow\uparrow}$ states and therefore couples the singlet and triplet states. In our experiment we measured the singlet-to-triplet exchange period to be greater than $10$ ms, which causes a negligible error in the scheme since the sideband coupling breaks the degeneracy of the singlet and triplet states. Fluctuations in the magnetic field also cause frequency offsets for the sideband and carrier drives by shifting the Zeeman splittings of the $^9$Be$^+$ energy levels.  The typical frequency offset for the sideband drive is small compared to $\Omega_s$ and therefore negligible.  The typical frequency offset for the carrier drive compared to $\Omega_s$ is more significant and leads to a decrease in the preparation since the $\ket{\uparrow}$ to $\ket{\rm a}$ transition will not be resonant. However, for our estimated magnetic-field fluctuations of approximately $10^{-7}$ T, we find an error for the singlet state of less than one percent for both the continuous and stepwise implementations using the numerical simulations.


\subsection{Heating Processes}
Heating processes compete with the sympathetic cooling and lead to a steady state with a thermal distribution. The largest heating process is caused by spontaneous emission from the $^{24}$Mg$^+$ ions during the application of the $^{24}$Mg$^+$ sideband Raman beams and repump light. Other smaller sources of heating are photon recoil due to spontaneous emission from the repumper and electric-field noise at the ions' positions (including the so-called anomalous heating \cite{Wineland1998}). These heating processes limit the lowest achievable $\bar{n}$ with sympathetic cooling for mode three to approximately $0.1$ for both the continuous and stepwise cases.  One error caused by these heating processes is the decrease in the singlet preparation rate as can be seen from Eq. \ref{EqPrep3}, which leads to an error of $0.02$ for the continuous case. However, if the only source of depumping from the singlet state is the inherent depumping (Eq. \ref{eq:gamma_inherent}), the fidelity for the continuous case can still be made to approach unity in the presence of heating by further increasing the sideband Rabi rate relative to other rates and leaving the interactions on for a longer duration.  Another source of error associated with the heating is due to the depumping process that results from unequal sideband Rabi rates on the $^{9}$Be$^+$ ions when the ions are not in the motional ground state, which leads to an error of $0.02$ for the continuous case as discussed in the previous section.

For the stepwise implementation, there is an additional error associated with the heating that is due to the $n$ dependence of $\Omega_{\rm eff}$, discussed in the main text, that leads to depumping from the $\ket{\rm S}$ state for $n \neq 0$. If we eliminate the heating processes in the numerical simulation of the stepwise implementation such that the ions are cooled to motional ground state the fidelity increases by approximately $0.04$. This error combines the effects of the decrease in preparation rate, the depumping due to sideband Rabi rate imbalance, and the additional depumping effect due to the $n$ dependence of $\Omega_{\rm eff}$.

\subsection{Off-Resonant Coupling}
Another potential source of error is off-resonant coupling of the $^{9}$Be$^+$ sideband beams to the qubit carrier transition or other motional mode sideband transitions. For our experimental parameters, the only significant coupling is that of the laser sideband to mode 4, which is detuned by approximately $\Delta\nu \approx 2 \pi \times 250$ kHz from the sideband laser drive. The Hamiltonian term for this coupling is $H_4 = {\Omega_{\rm s} \frac{\eta_4}{\eta_3}(\ket{\uparrow}_{1} \bra{\downarrow} -\ket{\uparrow}_{2} \bra{\downarrow})c^{+}e^{-i\delta t}+h.c.}$, where the minus sign occurs because the two $^{9}$Be$^+$ ions oscillate out of phase for mode 4. This couples $\ket{\rm S} \leftrightarrow \ket{\uparrow \uparrow} \ket{1}_4$, where the motional excitation is in the fourth mode. Cooling of this mode with a rate $\kappa_4 \approx 0.8$ kHz leads to an effective loss process from $\ket{\rm S}$ to $\ket{\uparrow \uparrow}$ at a rate
\begin{align}
\kappa^-_4 \approx \frac{2 \kappa_4 (\Omega_{\rm s} \frac{\eta_4}{\eta_3})^2}{\Delta \nu^2}.
\end{align}
The error due to this off-resonant coupling is estimated from the simulations to be $0.008$ for the continuous ($0.007$ when using Eq. (\ref{EqSteady})) and $0.023$ for the stepwise experiments. The strength of the off-resonant coupling could be reduced by using a better isolated motional mode frequency.
\begin{figure}
  \centering
  \includegraphics[width=8.6cm]{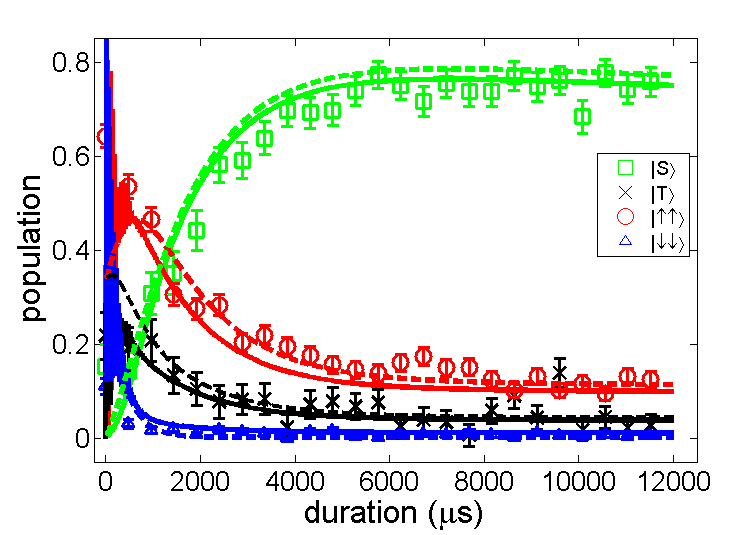}
  \caption{\textbf{Steady State Entanglement Data and Theory} We plot the dynamics of the ground state populations obtained by solving the rate equations (dashed lines) and the master equation (solid lines) together with the experimental data (symbols). While the rate equations do not capture the fast oscillations in the beginning, they agree well with the dynamics of the master equation and the experimental data for longer durations.}
\label{fig:curve}
\end{figure}

\subsection{Summary and Results}

In summary, we have derived the preparation rate $\gamma_+$ given in Eq. (\ref{EqPrep3}), the reshuffling rate $\kappa_{\rm res}$ in Eq. (\ref{EqRes}) and the loss rates
\begin{align}
\gamma^-_{\uparrow \uparrow} &= \frac{\gamma^-_{\rm inh} \gamma_{\rm \uparrow a}}{\gamma_{\rm \uparrow a} + \gamma_{\rm \downarrow a}} + \Gamma^-_{\uparrow \uparrow} + \kappa^-_r + \kappa^-_4\\
\gamma^-_{\rm T} &=  \frac{\gamma^-_{\rm inh} \gamma_{\rm \downarrow a}}{2 (\gamma_{\rm \downarrow a} + \gamma_{\rm \uparrow a})} + \Gamma^-_{\rm T} \\
\gamma^-_{\downarrow \downarrow} &= \Gamma^-_{\downarrow \downarrow} .
\end{align}
Using these rates we can model the experimental results by solving the coupled rate equations given by Eqs. (\ref{EqRate1})-(\ref{EqRate4}).
In Fig. \ref{fig:curve} we plot the evolution of the ground states that are obtained using the experimental parameters to calculate the rates derived in this section. In total, we find for the continuous implementation an error of about $0.23$ from the rate equation model, i.e., either from the steady-state fidelity in Eq. (\ref{EqSteady}) or from the simulation of Eqs. (\ref{EqRate1})-(\ref{EqRate4}). This is in good agreement with the value $0.24$ obtained from the simulation of the master equation and the experimental results.

Reaching higher-fidelity maximally entangled states with this scheme should be possible if spontaneous emission rates and imbalances of the sideband coupling on the qubits can be reduced. As an example, if the Raman detuning is increased to 1.5 THz, which reduces the spontaneous emission error by approximately a factor of 23 compared to the experiments presented here, and the sideband coupling imbalance is eliminated, the maximum achievable fidelity would be approximately $0.97$ using the continuous implementation. Here we have kept the same heating rate but assumed that errors due to off-resonant coupling have also been eliminated. To achieve the same sideband Rabi rate at this detuning, the laser intensity would need to be increased by a factor of 20.  Implementing the sideband coupling with microwaves would eliminate both spontaneous emission and unequal sideband Rabi rates and may therefore be a possible future approach to achieve high-fidelity entangled states with this dissipative scheme if other issues with this approach, such as high anomalous background heating, can be addressed \cite{Hite2012}.



%
%

\end{document}